\documentclass[journal,twocolumn]{IEEEtran}

\usepackage{blindtext, graphicx}
\usepackage{listings}
\usepackage{graphicx}
\usepackage[font=small]{caption}

\usepackage{color}
\usepackage{amsmath,bm}
\usepackage{amsmath}
\usepackage{amssymb}
\usepackage{algorithm}
\usepackage{algpseudocode}

\usepackage[utf8]{inputenc}

\makeatletter
\let\NAT@parse\undefined
\makeatother
\usepackage{hyperref}

\title{Optimal Adaptive Prediction Intervals for Electricity Load Forecasting in Distribution Systems via Reinforcement Learning}

\author{
\IEEEauthorblockN{Yufan Zhang, \textit{Student Member, IEEE},  Honglin Wen, \textit{Student Member, IEEE}, \\Qiuwei Wu, \textit{Senior Member, IEEE}, and Qian Ai, \textit{Senior Member, IEEE}}

\vspace{-2em}
\thanks{Yufan Zhang, Honglin Wen, and Qian Ai are with Department of Electrical Engineering, Shanghai Jiao Tong University, Shanghai 200240, China.\\
Qiuwei Wu is with Tsinghua-Berkeley Shenzhen Institute,Tsinghua Shenzhen International Graduate School, Tsinghua University, Shenzhen 518055, China.\\
Corresponding author: Qiuwei Wu (e-mail: qiuwu@sz.tsinghua.edu.cn).
}
}

\begin{document}

\maketitle
\thispagestyle{empty}
\pagestyle{empty}

\begin{abstract}

Prediction intervals (PIs) offer an effective tool for quantifying uncertainty of loads in distribution systems. The traditional central PIs cannot adapt well to skewed distributions, and their offline training fashion is vulnerable to the unforeseen change in future load patterns. Therefore, we propose an optimal PI estimation approach, which is online and adaptive to different data distributions by adaptively determining symmetric or asymmetric probability proportion pairs for quantiles of PIs’ bounds. It relies on the online learning ability of reinforcement learning (RL) to integrate the two online tasks, i.e., the adaptive selection of probability proportion pairs and quantile predictions, both of which are modeled by neural networks. As such, the quality of quantiles-formed PI can guide the selection process of optimal probability proportion pairs, which forms a closed loop to improve PIs’ quality. Furthermore, to improve the learning efficiency of quantile forecasts, a prioritized experience replay (PER) strategy is proposed for online quantile regression processes. Case studies on both load and net load demonstrate that the proposed method can better adapt to data distribution compared with online central PIs method. Compared with offline-trained methods, it obtains PIs with better quality and is more robust against concept drift.

Keywords: Prediction interval; Online load forecasting; Reinforcement learning; Uncertainty.

\end{abstract}

\section{Introduction}
Load forecasting plays a pivotal role in distribution systems operation, such as congestion management \cite{1} and energy transaction \cite{2}. However, it is widely acknowledged that accurate short-term load forecast (STLF) in distribution systems is challenging due to the uncertain characteristics of loads. Moreover, with the increasing penetration of renewable energy resources (RESs) in distribution systems, load measurements are blended with generation of RESs (and thus referred to as net loads), which further complicates the forecasting task. Although efforts have been devoted to accurate forecasts by focusing on both features \cite{3} and algorithms \cite{4}, point forecasts still cannot accommodate the inherent uncertainty of loads. The recent decade has witnessed the trend in load forecasting community, i.e., transition from point forecasts to probabilistic forecasts; see recent review \cite{5}.

In general, probabilistic short-term load forecasting (PSTLF) is to communicate the probability of load at future time, given information up to current time \cite{6}. It is usually developed in forms of quantile \cite{7}, prediction interval (PI) \cite{8}, and probability density \cite{9}. A PI is specified by two bounds and a nominal coverage probability (NCP) $(1-\beta)\times100\%$ , which indicates the probability that future observations fall in the interval. It has been successfully employed in the decision-making of distribution systems under uncertainty. For instance, PI delivers an uncertainty set for robust optimization (RO), such that RO produces a decision reliable for any realization in the set. In \cite{10}, multiple PIs with different NCP levels were formulated, so that RO was performed for the multiple PIs-constrained dispatch model, which reduced the conservatism of RO. Besides, PI offers information about uncertainty for interval optimization. In \cite{11}, PI was used to describe prediction uncertainty for the economic dispatch of virtual power plant, which was then converted into a deterministic problem. 

Traditionally, parametric approach with distributional assumption \cite{12}, residual resampling \cite{13}, and quantile regression (QR) \cite{14} are three popular approaches to constructing PIs. The parametric approach relies on distributional assumption, the shape parameters of which are estimated by data-driven methods. The delta, Bayesian, and mean-variance methods have been proposed \cite{12}, which assume that forecast errors have a Gaussian distribution. As residual resampling and QR are distribution-free, they are also referred to nonparametric approaches, whose successful applications in PI estimation can be traced back to \cite{14}. Resampling methods construct PI through sampling from the empirical distribution of point forecast errors. Ref. \cite{13} proposed a two-stage bootstrap sampling method for PSTLF. The first stage used bootstrap to characterize uncertainties from forecast models, and the second stage applied bootstrap to resample from the regression errors. Although the resampling methods are distribution-free, their performances are restricted by the point forecast methods relying on. QR directly estimates quantiles for loads, and uses two corresponding quantiles as the lower and upper bounds of PI. Ref. \cite{Cost-Oriented} proposed a cost-oriented PI method, where the quantile proportion pair of PI was selected to minimize operational cost. It is easy to combine QR with machine learning methods, for instance, extreme learning machine (ELM) \cite{15}, gradient boosting regression tree \cite{16}, and neural basis expansion model \cite{8}. A pinball-loss guided QR for load responsiveness estimation was proposed in \cite{17}, which was solved in a fully-distributed manner. In \cite{CQ}, an ensemble method for combining QR models was proposed, which was designed as a linear programming problem that minimized the pinball loss. Although QR method is distribution-free and easy to use, it faces the problem, i.e., \textit{how to determine the probability proportions for PI’s lower and upper bounds}. A typical approach is using the symmetric probability proportions. Such PI formed by $q^{\beta / 2}$ and $q^{1-\beta / 2}$ is referred to as a central prediction interval (CPI). Although these works have enriched tools for PSTLF and offered complementary understanding, recent studies reveal that CPIs are usually unnecessarily wide \cite{18}.

Therefore, increasing works have advocated the optimal prediction interval (OPI) that aims to minimize PI width under the constraint of coverage. One of the most popular approaches is to estimate the PI’s bounds directly, with the objective accounting for both reliability and sharpness, which is therefore free of quantile proportions selection. Ref. \cite{19} developed a neural network (NN) model to output the bounds. The NN was trained by minimizing the PI-based objective function. However, for the NN-based method, the PI-based objective function needs to be carefully designed, such that it is differentiable for NN to learn. To solve this problem, \cite{cpi} proposed a relaxed surrogate to the original non-differentiable objective, making gradient descent a viable solution method. However, the loss functions in these two works are not proper metrics in the view of statistics, which translates into saying that the quality of the forecasts issued by these two methods may be a concern in the term of proper statistical scores. Ref. \cite{20} proposed a hybrid method of OPI based on ELM and particle swarm optimization. The multi-objective optimization for the ELM’s output weights determination was developed. Although there is no differentiable requirement for the PI-based objective by using particle swarm optimization, the multi-objective optimization involving the soft constraints is hard to obtain the Pareto optimal, which hinders accurate estimation of the PI. Another approach is to optimize the selection of quantiles’ probability proportions. For instance, \cite{18} employed an ELM to generate nonlinear features, and mixed-integer linear programming (MILP) was formulated to obtain the OPI. The ELM does not involve backpropagations, which means the feature learning stage is inefficient and leads to the obstacles in the later estimation stage. Therefore, it remains an open issue to develop OPI.

Another limitation of the aforementioned works stem from offline-trained algorithms. Their frameworks contain offline training and subsequent online forecasting. However, offline-trained algorithms rely on the common assumption that training and test data are drawn from the same distribution, also referred to as independently identically distribution. In reality, concept drift often occurs \cite{21}. In other words, the distribution of test data differs from that of training data. Usually, offline models cannot adapt well to the changing distribution. Moreover, to retrain offline models, it usually requires abundant recollected samples, and takes expensive computation. Indeed, online learning is an appealing solution to tackle this issue \cite{22}. In this context, the model iteratively predicts a value, and then receives a cost according to an adversarially chosen function. The parameters of the model are updated in an online fashion, with only newly observed sample or a batch of samples. An online ensemble learning approach for load forecasting was proposed in \cite{23}. Using the passive aggressive regression method, the convex optimization problem was formulated for the new observation to update the ensemble weights. However, as this approach is grounded in convex optimization, it is not suitable for the parameter estimation of nonconvex models. Using recurrent neural network (RNN) for online prediction, \cite{24} updated RNN's parameters by a part of data batch sampled from the buffer. However, since the samples stored in the buffer are assumed to have the equal significance for model updating, their different contributions to improving the forecast performance can hardly be distinguished. 

Indeed, the OPI problem can be decoupled into two correlated sub-problems, i.e., probability proportion selections and corresponding quantile forecasts through QR models fitting. As such, the tasks of proportion selections and QR fitting can be fulfilled separately, which frees the limitation imposed on QR models in \cite{18}. And the task of proportion selections is nontrivial and cannot be searched in an exhaustive manner, as probability proportions are continuous. Instead, it can be modeled as a decision-making problem that an agent decides the proportions given some contexts, which can be further naturally considered in an online setting. In this paper, the OPI problem at each time-step is formulated as a bilevel program, where the upper level performs the quantile proportion selection, and given the selected proportion, the lower level predicts quantiles for PI construction. Reinforcement learning (RL) is leveraged to solve the bilevel program in an online fashion, which integrates upper and lower decision tasks in a closed-loop manner. Specifically, the upper-level task is modeled by the decision making of the agent, whereas the lower-level task is solved in the environment. RL has strong theoretical background rooted in Markov Decision Process (MDP). And the bounded reward value guarantees the online iterations converging to the optimal value \cite{25}. Recent years have witnessed its successful applications, such as in the online energy management \cite{26,27}. Therefore, this paper proposes an online OPI forecasting method that formulates PIs in an adaptive manner. In general, it is a closed-loop feedback framework working toward the quality improvement of estimated PIs. Concretely, the closed-loop framework is established based on the value-based RL algorithm, where the agent is modeled by a neural network. At each time-step, the agent receives an input state, and adaptively determines the symmetric or asymmetric quantile proportion pairs for PI as corresponding action. Based on the action, the environment estimates PI’s bounds via the online QR models. Then, the parameters of online QR models are updated through gradient descent optimization by a batch of data from the buffer. In particular, an online quantile forecasting method with prioritized experience replay (PER) \cite{28} is proposed. Unlike treating samples in the buffer equally, here the data is assigned with different probability of being sampled to update the model parameters, which is derived by their importance for improving the QR’s performance. After the quantile predictions, the overall quality of the formed PI is used as the feedback to guide the agent in return. Compared with the existing studies, the main contributions of the paper are summarized as follows: 

(1) The proposal of a new approach for OPI, which is adaptive to the underlying data distribution, and uncovers the relationship between optimal quantile proportion selection and quantile forecasting for OPI estimation;

(2) A RL-based online OPI prediction framework, which integrates the quantile proportion selection and quantile estimations for PI’s lower and upper bounds (LUB). As such, the two tasks work in a closed-loop manner toward PI quality improvement. Also, under such a framework, a large spectrum of QR models are applicable to the quantile forecasting task;

(3) An online quantile forecasting approach with the PER strategy, which improves the data efficiency and QR models’ prediction quality by learning more frequently on the prioritized samples.

The remainder of this paper is organized as follows. Section \uppercase\expandafter{\romannumeral2} introduces the definition of the proposed problem and the overall framework. Section \uppercase\expandafter{\romannumeral3} proposes the online OPI algorithm. Results are discussed and evaluated in Section \uppercase\expandafter{\romannumeral4} followed by the conclusions.

\section{Problem Statement}

In this section, we formulate the OPI problem mathematically in subsection \emph {A}, and present the overall framework in subsection \textit{B}. 

\subsection{Problem Definition}

Let $y_t$ denote the load at time $t$, which is a realization of the corresponding random variable $Y_t$. The probabilistic load forecasting with lead time $k$ translates to communicate the density function $f_{t+k|t}(Y_{t+k}|\Omega_t)$ and cumulative distribution function $F_{t+k|t}(Y_{t+k}|\Omega_t)$, where $\Omega_t=\left\{Y_{t+h-1},...,Y_{t}\right\}$ is the information set up to time $t$. Specifically, a quantile $q_{t+k}^{\alpha_{t+k}}$ is defined as $F_{t+k|t}^{-1}(\alpha_{t+k})$ where the quantile’s probability proportion $\alpha_{t+k}\in	\left[ 0,1 \right]$.  Therefore, a PI with NCP $(1-\beta)\times100\%$ can be developed as: 
$$ \left[ \hat{q}_{t+k}^{\alpha_{t+k}},\hat{q}_{t+k}^{\alpha_{t+k}+1-\beta} \right]. \eqno{(1)}$$

Here the quantiles are the outputs of QR models
\begin{equation}
\begin{split}
&\hat{q}_{t+k}^{\alpha_{t+k}}=f^{\alpha_{t+k}}(\bm{s}_t;\hat{\bm{W}}^{\alpha_{t+k}}_t)\\
&\hat{q}_{t+k}^{\alpha_{t+k}+1-\beta}=f^{\alpha_{t+k}+1-\beta}(\bm{s}_t;\hat{\bm{W}}^{\alpha_{t+k}+1-\beta}_t),
\end{split}\tag{2}
\end{equation}
where the vector $\bm{s}_t$ is the realization of $\Omega_t$ which is composed of the $h$ previous values, and $\hat{\bm{W}}^{\alpha_{t+k}}_t$, $\hat{\bm{W}}^{\alpha_{t+k}+1-\beta}_t$ are QR models' parameters for the estimations of LUB’s quantiles. The estimated parameters of QR models can be obtained by solving the optimization problem of minimizing a pinball loss

\begin{equation}
\begin{split}
&\hat{\bm{W}}_t^{\alpha_{t+k}}=\mathop{\arg\min}_{\bm{W}_t^{\alpha_{t+k}}}\mathbb{E}_{F_{t+k|t}}L^{\alpha_{t+k}}(Y_{t+k},q_{t+k}^{\alpha_{t+k}})\\
& \qquad \qquad s.t. q_{t+k}^{\alpha_{t+k}}=f^{\alpha_{t+k}}(\bm{s}_t;\bm{W}^{\alpha_{t+k}}_t)\\
&\hat{\bm{W}}_t^{\alpha_{t+k}+1-\beta}=\mathop{\arg\min}_{\bm{W}_t^{\alpha_{t+k}+1-\beta}}\mathbb{E}_{F_{t+k|t}}L^{\alpha_{t+k}+1-\beta}(Y_{t+k},q_{t+k}^{\alpha_{t+k}+1-\beta})\\
& \qquad \qquad s.t. q_{t+k}^{\alpha_{t+k}+1-\beta}=f^{\alpha_{t+k}+1-\beta}(\bm{s}_t;\bm{W}^{\alpha_{t+k}+1-\beta}_t),
\end{split}\tag{3}
\end{equation}
where $L^\alpha$, $\alpha \in \{\alpha_{t+k},\alpha_{t+k}+1-\beta\}$ is the pinball loss, which is defined as
\begin{equation}
\begin{split}
L^\alpha(y_{t+k},q_{t+k}^\alpha)= 
(1-\alpha)(q_{t+k}^\alpha-y_{t+k})I_{\left\{q_{t+k}^\alpha-y_{t+k} \geq 0 \right\} }\\ +\alpha(y_{t+k}-q_{t+k}^\alpha)I_{\left\{q_{t+k}^\alpha-y_{t+k} \textless 0 \right\}},
\end{split}\tag{4}
\end{equation}
where $I_{\left\{ . \right\}}$ is the indicator function.

For symmetric distributions like Gaussian distribution, $\alpha_{t+k}$ is often determined as $\beta/2$, for any time indiscriminately. However, for asymmetric distributions such as log-normal distribution, Beta distribution, such CPIs cannot adapt well to the real distribution. Indeed, conditioned on a certain data distribution $F_{t+k|t}(Y_{t+k}|\Omega_t)$, there exists an OPI $ \left[ \hat{q}_{t+k}^{\alpha_{t+k}^*},\hat{q}_{t+k}^{\alpha_{t+k}^*+1-\beta} \right]$ with optimal proportions $\alpha_{t+k}^*,\alpha_{t+k}^*+1-\beta$, whose width is smaller than that of CPI with the same NCP. The illustration of the differences between CPI and OPI under a skewed probability distribution is shown in Fig. 1. The shaded parts have the same area under the density curve. Since the probability density function is with positive skewness, the width of OPI is smaller than that of CPI. 

\begin{figure}[h]
  \centering
  \includegraphics[scale=0.4]{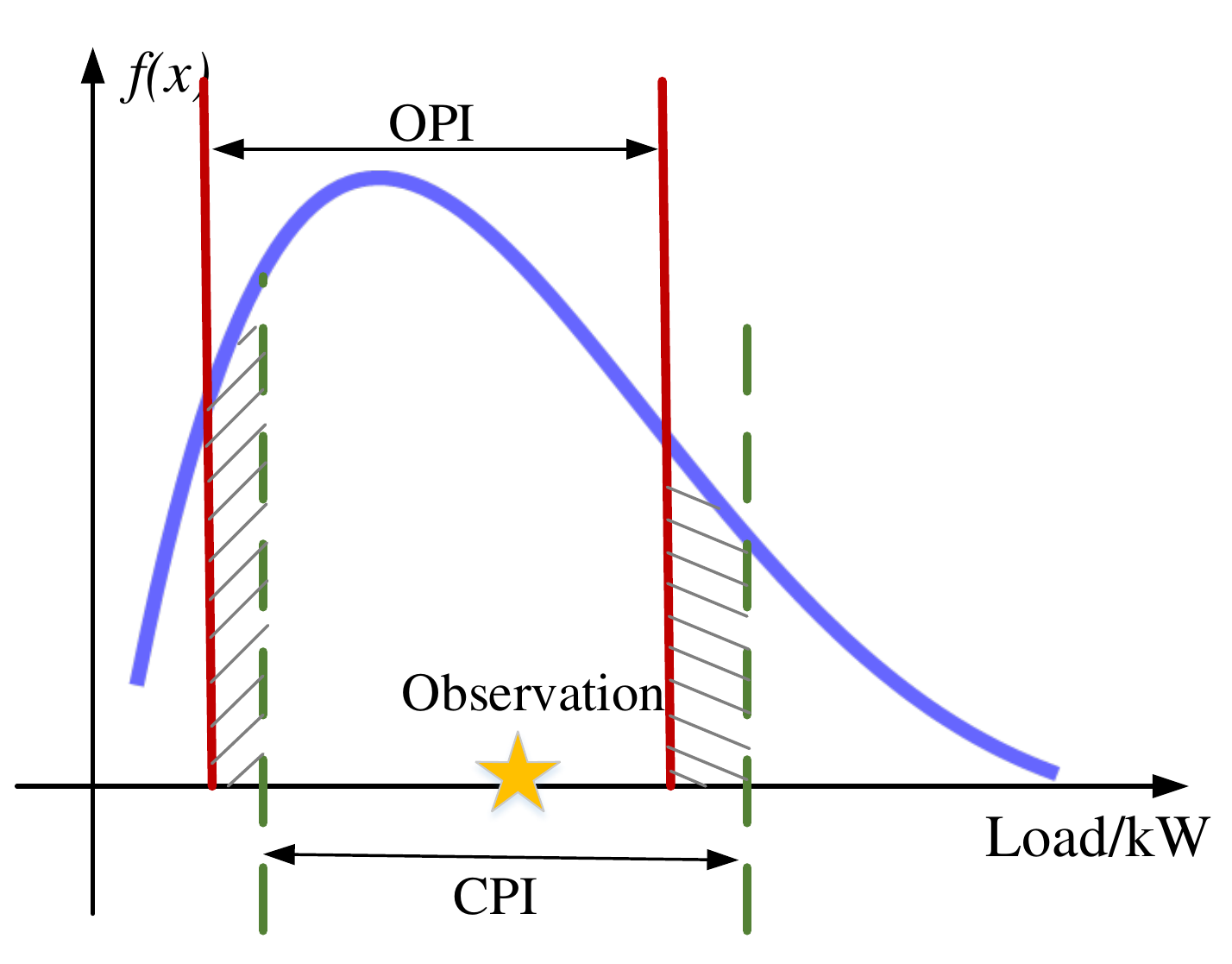}\\
  \caption{Illustration of CPI and OPI under a skewed probability density function. }
\end{figure}

In this paper, instead of treating the quantile proportion as the predetermined constant, the problem of OPIs formulation involves selection of quantile proportions and corresponding quantile estimations. That is, we wish to find the optimal proportion pair at time $t$, such that the Winkler score $w(y_{t+k},\hat{q}_{t+k}^{\alpha_{t+k}^*},\hat{q}_{t+k}^{\alpha_{t+k}^*+1-\beta})$ that measures the overall quality of PI is minimized. The selected proportion pair at time $t$ is then used to specify corresponding QR models. Therefore, the OPI problem at time $t$ can be formulated as a bilevel programming problem, 
\begin{equation}
\begin{split}
&\mathop{\min}_{\bm{\theta}_t}w(y_{t+k},\hat{q}_{t+k}^{\alpha_{t+k}^*},\hat{q}_{t+k}^{\alpha_{t+k}^*+1-\beta})\\
& s.t. \\
&\alpha_{t+k}^*=f(\bm{s}_t;\boldsymbol{\theta}_t)\\
&\hat{q}_{t+k}^{\alpha_{t+k}^*}=f^{\alpha_{t+k}^*}(\bm{s}_t;\hat{\bm{W}}^{\alpha_{t+k}^*}_t)\\
&\hat{q}_{t+k}^{\alpha_{t+k}^*+1-\beta}=f^{\alpha_{t+k}^*+1-\beta}(\bm{s}_t;\hat{\bm{W}}^{\alpha_{t+k}^*+1-\beta}_t)\\
&\hat{\bm{W}}_t^{\alpha_{t+k}^*}=\mathop{\arg\min}_{\bm{W}_t^{\alpha_{t+k}^*}}\mathbb{E}_{F_{t+k|t}}L^{\alpha_{t+k}^*}(Y_{t+k},q_{t+k}^{\alpha_{t+k}^*})\\
& \qquad \qquad s.t. q_{t+k}^{\alpha_{t+k}^*}=f^{\alpha_{t+k}^*}(\bm{s}_t;\bm{W}^{\alpha_{t+k}^*}_t)\\
&\hat{\bm{W}}_t^{\alpha_{t+k}^*+1-\beta}=\mathop{\arg\min}_{\bm{W}_t^{\alpha_{t+k}^*+1-\beta}}\mathbb{E}_{F_{t+k|t}}L^{\alpha_{t+k}^*+1-\beta}(Y_{t+k},q_{t+k}^{\alpha_{t+k}^*+1-\beta})\\
& \qquad \qquad s.t. q_{t+k}^{\alpha_{t+k}^*+1-\beta}=f^{\alpha_{t+k}^*+1-\beta}(\bm{s}_t;\bm{W}^{\alpha_{t+k}^*+1-\beta}_t).
\end{split}\tag{5}
\end{equation}

As each two adjacent states are auto-correlated, the state at time $t$ transits to the state at time $t+1$ according to the state transition function $S^M(.)$, i.e., $\bm{s}_{t+1}=S^M(\bm{s}_t)$. It involves a similar problem that minimizes the Winkler score at the time $t+1$.

In (5), the upper level performs the probability proportion selection by learning a mapping $f(\bm{s}_t;\bm{\theta}_t)$ with parameters $\bm{\theta}_t$. The lower-level problem performs the quantile forecasting. In general, the two problems in two levels are interdependent. Usually, the QR model is highly nonconvex and the bilevel problem in (5) is difficult to solve. However, in some special case where the lower-level QR model is linear, the bilevel problem can be transformed to a single level mathematical programming problem. For the general problem, we propose to solve it in an iterative manner via RL framework. Also, as random variable’s cumulative distribution function $F_{t+k|t}$ changes with on-going input data, the upper- and lower-level optimization problem changes from time to time. Therefore, the selection of optimal proportion should be updated with time. Besides, the estimation of QR models should also be updated with time, due to the nonstationary characteristics of the time series. Therefore, at time $t$, we optimize over the probability proportion and corresponding quantiles.

\subsection{The Proposed Framework}
In this paper, RL serves as the basic closed-loop framework for integrating the aforenamed tasks. Specifically, to issue the quantile forecasts for load at time $t$ , one uses previous values $\left\{y_{t-h},...,y_{t-1}\right\}$ as input features and estimates $\hat{q}_{t-1+k}^{\alpha_{t-1+k}^*}$, $\hat{q}_{t-1+k}^{\alpha_{t-1+k}^*+1-\beta}$ via parameters $\hat{\bm{W}}^{\alpha_{t-1+k}^*}_{t-1}$, $\hat{\bm{W}}^{\alpha_{t-1+k}^*+1-\beta}_{t-1}$. Then after $y_t$ is revealed, the parameter of the corresponding QR models $f^{\alpha_{t-1+k}^*}(.)$, $f^{\alpha_{t-1+k}^*+1-\beta}(.)$ are updated by gradient descent at the time $t$ as: $\hat{\bm{W}}^{\alpha_{t-1+k}^*}_{t}$, $\hat{\bm{W}}^{\alpha_{t-1+k}^*+1-\beta}_{t}$, while the parameters of the remaining QR models at the time $t$ remain the same as that of their parameters at the time $t-1$. Then, after $\alpha_{t+k}^*$ at the time $t$ is revealed, quantiles $\hat{q}_{t+k}^{\alpha_{t+k}^*}$, $\hat{q}_{t+k}^{\alpha_{t+k}^*+1-\beta}$ are estimated by $f^{\alpha_{t+k}^*}(. ;\hat{\bm{W}}^{\alpha_{t+k}^*}_{t})$, $f^{\alpha_{t+k}^*+1-\beta}(. ;\hat{\bm{W}}^{\alpha_{t+k}^*+1-\beta}_{t})$. For simplicity of notations, $\left\{y_{t-h},...,y_{t-1}\right\}$ is denoted as $\bm{x}_t$. The whole framework is illustrated in Fig. 2.

\begin{figure}[h]
  \centering
  \includegraphics[scale=0.4]{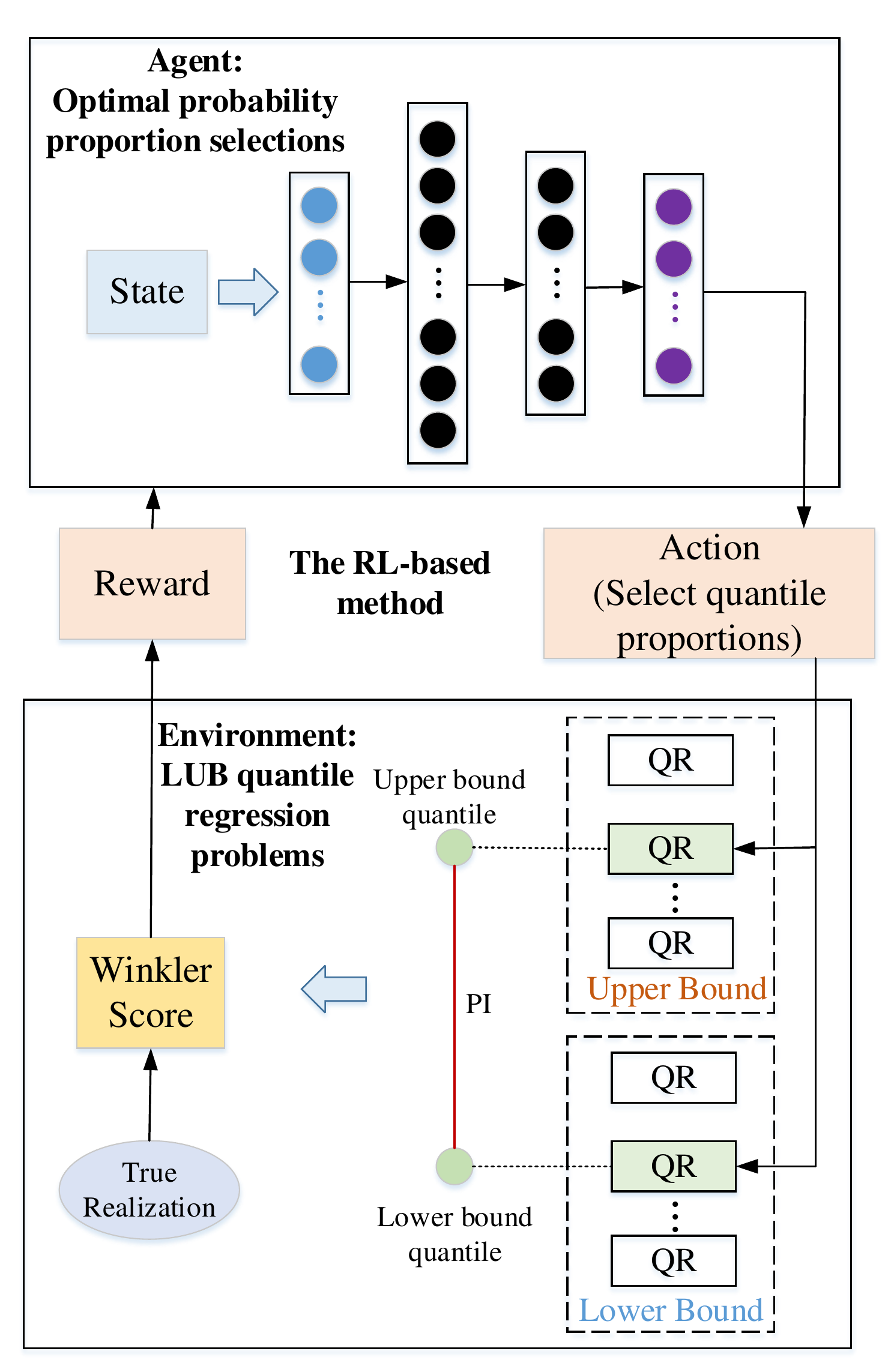}\\
  \caption{Illustration of proposed RL-based OPI framework. }
\end{figure}

As shown, an agent, modeled by NN, is responsible for selecting the optimal quantile proportion $\alpha_{t-1+k}^*$ for lower bound’s quantile under the current state/context $\bm{x}_t$. As the probability proportion is continuous in the range of $(0,\beta)$ , the optimal proportion cannot be found by exhaustive enumeration, which leads to the adoption of RL method. And corresponding QR models with different quantile proportions are in the environment for predicting the LUB’s quantiles for PIs. Specifically, the states of the agent are also the input features fed into QR models. And the quality of the constructed PIs is used as the feedback reward to link the two tasks. Since we focus on online estimation, the lead time $k$ equals one in the following discussion. We detail the RL formulation of the online OPIs construction problem. The key elements of which are defined as follows:

\textit{State}: For state at the time $t$, the QR’s input feature $\bm{x}_t$ is also used as the state for the RL’s agent.

\textit{State transition function}:
To issue the quantile forecasts for load at time $t$, $\bm{x}_{t}=\{y_{t-h},...,y_{t-1}\}$ is the state of the agent as well as input features of the QR model. Likewise, to issue the quantile forecasts for load at time $t+1$, the state of the agent is $\bm{x}_{t+1}=\{y_{t-h+1},...,y_{t}\}$. The state transition from $\bm{x}_{t}$ to $\bm{x}_{t+1}$ in this problem is a mapping of continuous variables and independent of the action. Since the state transition of high-dimensional continuous states is difficult to model, it is hard to solve the formulated DRL problem via a model-based RL method that involves learning a dynamic model. Therefore, we decide to model the abovementioned RL problem in a model-free approach, rather than a traditional model-based RL approach.

\textit{Action}: To choose optimal probability proportion from the continuous range of $(0,\beta)$, the probability proportion is transformed into discrete variables. Concretely, the discrete action space $\mathcal{A}$ is a set of probability proportions with the size of $|\mathcal{A}|$, which is equal to the number of quantile proportions. With the increasing number of actions, the range of $(0,\beta)$ is modeled more precisely for the probability proportion of lower bound’s quantile. Given the certain NCP, the action space is:
$$\left\{\frac{i\cdot\beta}{|\mathcal{A}|+1}\right\}^{|\mathcal{A}|}_{i=1},\eqno{(6)}$$
where $|\mathcal{A}|=2^n-1, n=1,2,...$ is the number of actions. So, at the time $t$, the optimal $\alpha^*_t$ is chosen from the set defined in (6). When $n=1$, the problem returns back to CPI. The illustration of the action space with the change of action numbers is shown in Fig. 3. The action space with the larger dimension contains all the actions in the previous action space with the smaller dimension. Although the number of actions has impact on the agent’s policy, as the difference between adjacent quantile proportions under relatively large action size is small, the impact of the action space’s size on the selection policy is marginal.

\textit{Remark 1}:  In the general solution of the model in (5), the optimal solution of the upper-level problem in (5) can be continuous probability proportion. However, we choose to approximate the continuous proportion by discrete action space out of two practical concerns:

Firstly, as we train QR models separately for each quantile proportion in the action space, with finite discrete probability proportions, the number of QR models is also finite. So, it is only required to optimize the parameters of limited number of QR models. In contrast, if probability proportions are modeled in a continuous action space, the number of QR models will be infinite, which considerably increases the computational burden. Secondly, since the adjacent probability proportions in the action space are close to each other, the discrete probability proportions are adequate for representing the continuous probability proportions in a range.

\begin{figure}[h]
  \centering
  \includegraphics[scale=0.5]{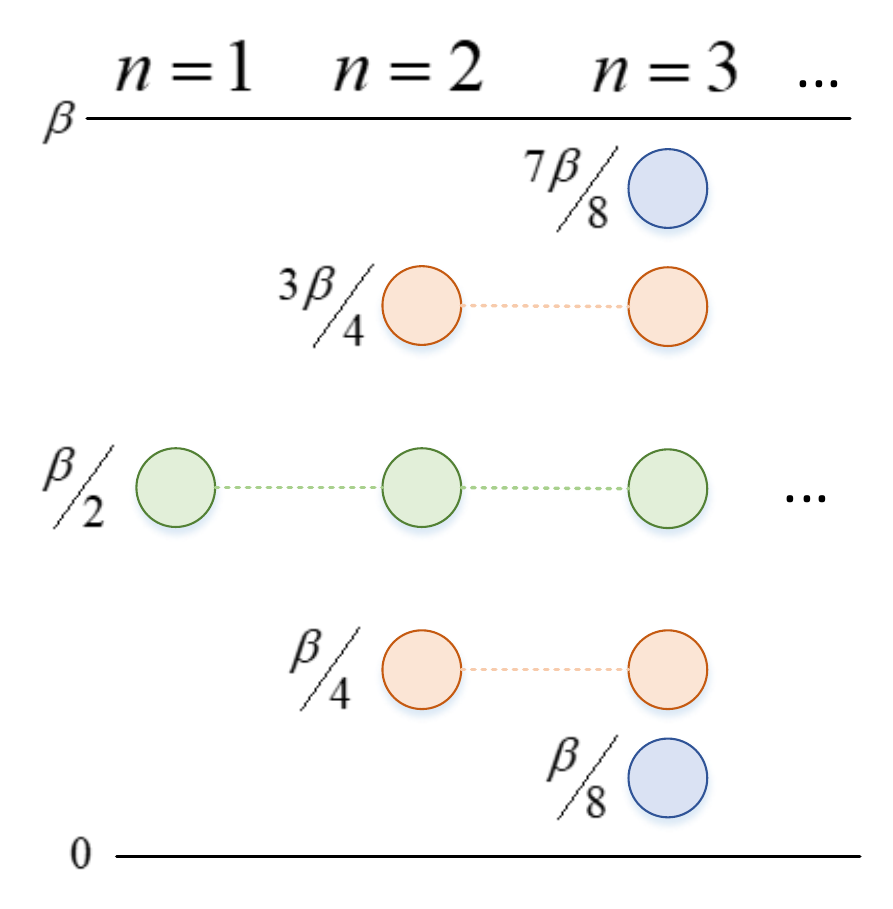}\\
  \caption{Illustration of action space. }
\end{figure}

\textit{Reward}: The Winkler score is a comprehensive evaluation index of PI quality, which measures both reliability and sharpness simultaneously. The smaller the Winkler score, the better the quality of the constructed PI. It is defined as:
$$
w_t=\left\{
\begin{aligned}
&\hat{q}_{t}^{\alpha_{t}^*+1-\beta}-\hat{q}_{t}^{\alpha_{t}^*},  \qquad\quad\quad\qquad \hat{q}_{t}^{\alpha_{t}^*}\leq y_t\leq \hat{q}_{t}^{\alpha_{t}^*+1-\beta} \\
&\hat{q}_{t}^{\alpha_{t}^*+1-\beta}-\hat{q}_{t}^{\alpha_{t}^*} + 2(\hat{q}_{t}^{\alpha_{t}^*}-y_t)/\beta,  \quad\quad\quad\quad y_t < \hat{q}_{t}^{\alpha_{t}^*} \\
&\hat{q}_{t}^{\alpha_{t}^*+1-\beta}-\hat{q}_{t}^{\alpha_{t}^*} + 2(y_t-\hat{q}_{t}^{\alpha_{t}^*+1-\beta})/\beta,  \hat{q}_{t}^{\alpha_{t}^*+1-\beta} < y_t
\end{aligned}
\right.\eqno{(7)}
$$

Therefore, the negative value of Winkler score for the constructed PI at time-step $t$ is used as the reward:
$$ r_t=-w_t.\eqno{(8)}$$

\textit{Agent}: The optimal probability proportion selection problem is solved by agent, which gradually learns how to choose the optimal probability proportion for the lower bound quantile to construct good quality PI.

\textit{Environment}: The LUB’ quantiles forecasting problem is in the environment, which is consisted of $2\cdot|\mathcal{A}|$ quantile predictors which are QR models. The online forecasting process is illustrated in the next Section and each predictor has its own model parameters and buffer.

\section{Online Optimal Construction of PI}

In this Section, a general form of online quantile forecasting with PER strategy is proposed in subsection \textit{A}. Based on it, the proposed online OPI method is presented in subsection \textit{B}.

\subsection{Online Quantile Forecasting with PER Strategy}
Contrary to generating the best quantile predictor on the entire training set at once, online QR updates the predictor for future load at each step. Therefore, in the case of the $\alpha_{th}$ quantile online forecasting, the QR model $f^{\alpha}(.;\bm{W}_t^\alpha)$ with parameters $\bm{W}_t^\alpha$ is updated in a recursive and adaptive manner for data in a sequential order. Here, we update the model using gradient descent recursively by a small batch of data. Specifically, the new data revealed is stored into the predictor’s buffer $D_t^\alpha$, and a small batch of feature and label pairs, which are referred as experiences, are sampled from the buffer to update the model parameters $\bm{W}_t^\alpha$. It has several advantages over the standard online stochastic gradient descent. First, the data efficiency is improved, because the historical data pairs can be potentially used in many updates. Second, since the parameter updating is based on a small batch of data not a single data, the parameters learning process can be more stable. 

Here, instead of sampling data from the buffer uniformly at random, the online quantile forecasting with PER strategy is proposed, where the data are sampled with priority. The priority of sampling experience $e_j=(\bm{x}_j,y_j)$ is defined as:
$$p^\alpha_j=L^\alpha(y_j,\hat{q}_j^\alpha),\eqno{(9)}$$
where $\hat{q}_j^\alpha=f^{\alpha}(\bm{x}_j;\hat{\bm{W}}_{t-1}^\alpha)$ is the estimated quantile and $y_j$ is the realization of the load. $L^\alpha(y_j,\hat{q}_j^\alpha)$ is the pinball loss of the $e_j$, and $p_j^\alpha$ is the priority. 

And for the new data, its priority is assigned with the maximum priority in the predictor’s buffer, thus favoring the new data during sampling later. 

The probability of the experience $e_j$ being sampled is defined based on the priority as follows:
$$P_j=(p_j^\alpha)^\sigma \big/ \sum_{k} (p_k^\alpha)^\sigma,\eqno{(10)}$$
where $\sigma$ controls the extent that the prioritization is used. When $\sigma=0$, it corresponds to the case of uniform sampling. Therefore, by assigning data with probability proportional to the priority, the experience with large priority is more likely to be drawn frequently for parameter updating. Therefore, with PER strategy, the sample with larger pinball loss and the new sample are more likely to be sampled for parameters updating. That is, the samples that are harder to learn and that bring new information are learned more frequently.

With PER strategy, the loss function $\ell(\hat{\bm{W}}^\alpha_{t-1})$ of a small batch of experiences with the size $B$ can be written as follows:
$$\ell(\hat{\bm{W}}^\alpha_{t-1})=\frac{1}{B}\sum_{j}w_j(p_j^\alpha)^2,\eqno{(11)}$$
$$w_j=(N\cdot P_j)^{-\rho} \big/ w^{max},\eqno{(12)}$$
where $w_j$ is the normalized importance-sampling (IS) weight, which corrects the bias introduced by prioritized sampling \cite{28}. Through the correction, (12) shows that the experience with larger probability being sampled is assigned with the smaller weight. $N$ is the number of experiences in the predictor’s buffer, and $w^{max}$ is the maximum IS weight. The parameter $\rho$ controls the extent of importance-sampling correction, which fully compensates for the priority sampling when $\rho=1$. When $\rho=0$, there is no correction. 

Then, the predictor’s parameters update is as follows:
$$\hat{\bm{W}}^\alpha_{t}=\hat{\bm{W}}^\alpha_{t-1}-\eta_\alpha\cdot\nabla_{\hat{\bm{W}}^\alpha_{t-1}}\ell(\hat{\bm{W}}^\alpha_{t-1}).\eqno{(13)}$$

Here, $\eta_\alpha$ is the learning rate of the QR model. The final online quantile forecasting method with PER strategy is outlined in Algorithm 1, and the output of Algorithm 1 is the predicted quantile at each time-step. The additional remarks are presented below.

\begin{algorithm}[h]
	\caption{Online Quantile Forecasting with PER Strategy}
	\label{alg1}
	\begin{algorithmic}[1]
	\Require{Batch size $B$, learning rate $\eta_\alpha$, parameters $\sigma$,$\rho$, the quantile’s probability proportion $\alpha$}
	\State{Initialize the quantile predictor $f^\alpha(.;\hat{\bm{W}}_0^\alpha)$ with random weights $\hat{\bm{W}}_0^\alpha$}
	\For{$t$=1,2,...}
	    \State{Given the input feature $\bm{x}_t$, predict the quantile $\hat{q}_t^\alpha=f^\alpha(\bm{x}_t;\hat{\bm{W}}_{t-1}^\alpha)$}
	    \State{Reveal true value $y_t$}
	    \State{Store the experience $e_t=(\bm{x}_t,y_t)$ in the predictor’s buffer $D_t^\alpha$ with maximal priority $p_t^\alpha=max_{i<t}p_i^\alpha$}
	    \If{the size of predictor’s buffer $|D_t^\alpha| \geq B$}
	        \For{$j=1,...,B$}
	            \State{Sample experience $e_j$ from $D_t^\alpha$ according to $P_j$ calculated by (10)}
	            \State{Compute IS weight by (12)}
	            \State{Update priority by (9)}
	       \EndFor
	       \State{Update the model parameters by (13)}
	   \EndIf
	   \State{Update the input feature $\bm{x}_{t+1}$ at the next time-step based on the true load value $y_t$}
	    
	 \EndFor

	\end{algorithmic}

\end{algorithm}

Line 3 – Line 5: The input feature $\bm{x}_t$ and target variable $y_t$ are the new data at the time-step $t$, which are stored into the predictor’s buffer $D_t^\alpha$ at each time-step. 

Line 6 – Line 13: By drawing samples from the predictor’s buffer $\left\{e_j\right\}_{j=1}^B\sim D_t^\alpha$ according to the probability, the predictor’s parameters are updated on the small batch of experiences. 

Line 14: After parameter updating, the feature at the next time-step is updated based on the true load value $y_t$ at time $t$ using the moving-split window method. Then, the predictor estimates the next time-step quantile based on the updated feature and model parameters.

\subsection{The Proposed Online OPI Method}
The proposed online OPI method is shown in Algorithm 2. The value-based RL method dueling deep Q-networks (DQN) is used. We note that the specific RL method is not the main focus of the work, and many other RL methods with discrete action space can also be applied. Dueling DQN \cite{30} divides the estimation of state-action value function $Q_\pi(\bm{x}_t,\bm{\alpha})$ under the policy $\pi$ into two streams: the respective estimations of state value $V_\pi(\bm{x}_t)$ and state-dependent action advantages $A_\pi(\bm{x}_t,\bm{\alpha})$. Here, $\bm{\alpha}$ is the action vector formed by the discrete proportions in the set defined in (6). Therefore, the state-action value for selecting any proportion in $\bm{\alpha}$ at time $t$ can be expressed as:
$$Q_\pi(\bm{x}_t,\alpha_t)=V_\pi(\bm{x}_t)+(A_\pi(\bm{x}_t,\alpha_t)-\frac{1}{|\mathcal{A}|}\sum_{\alpha_t^\prime \in \mathcal{A}} A_\pi(\bm{x}_t,\alpha_t^\prime)),\eqno{(14)}$$
where $\alpha_t$ is the selected proportion. 

Based on the state-action value (14), to balance the exploration and exploitation, the action is chosen in a $\epsilon$-greedy manner.

Since the forecast errors measured by Winkler score at different time (for instance $t$ and $t+v$, where $v$ is a positive integer) are autocorrelated \cite{correrror}, the reward at time $t$ is correlated with the reward at time $t+v$. So the agent needs to be farsighted to consider the future reward at the current time-step, which matches the delayed reward feature of RL. Therefore, it is natural to model the state-action value at time $t$ as the expected discounted reward function:
$$Q_\pi(\bm{x}_t,\alpha_t)=E_\pi[r_t+\gamma r_{t+1} + \gamma^2 r_{t+2} + ...|S_t=\bm{x}_t,A_t=\alpha_t],\eqno{(15)}$$
where $\gamma$ is the discount factor.

Using the Bellman equation, the current state-action value is decomposed into immediate reward plus discounted state-action value of successor state. Therefore, the agent of dueling DQN performs the minimization of temporal difference (TD) error: 
$$L=\mathbb{E}[(r_t+\gamma \mathop{max}\limits_{\alpha_{t+1}}Q_{\pi^\prime}(\bm{x}_{t+1},\alpha_{t+1};\bm{\theta}^-)-Q_\pi(\bm{x}_{t},\alpha_{t};\bm{\theta}))^2],\eqno{(16)}$$
where $Q_{\pi^\prime}(\bm{x}_{t+1},\alpha_{t+1};\bm{\theta}^-)$ is estimated by the target network, and $Q_\pi(\bm{x}_{t},\alpha_{t};\bm{\theta})$ is estimated by the local network, and $\bm{\theta}^-,\bm{\theta}$ are the parameters of target and local networks respectively. Based on (16), the local network is updated by:
\begin{equation}
\begin{split}
&\frac{\partial L}{\partial \bm{\theta}}=\eta_Q\cdot  \mathbb{E}[(r_t+\gamma \mathop{max}\limits_{\alpha_{t+1}}Q_{\pi^\prime}(\bm{x}_{t+1},\alpha_{t+1};\bm{\theta}^-)-\\ &Q_\pi(\bm{x}_{t},\alpha_{t};\bm{\theta})) \frac{\partial Q_\pi(\bm{x}_{t},\alpha_{t};\bm{\theta})}{\partial \bm{\theta}}],
\end{split}\tag{17}
\end{equation}
where $\eta_Q$ is the agent's learning rate. 

Then, the target network’s parameters are updated by:
$$\bm{\theta}\cdot \tau+(1-\tau)\cdot \bm{\theta}^-\rightarrow \bm{\theta}^-,\eqno{(18)}$$
where $\tau$ is the soft update parameter.

\begin{algorithm}[h]
	\caption{Online OPI Method}
	\label{alg1}
	\begin{algorithmic}[1]
	\Require{Batch size $B$, learning rates $\eta_\alpha$,$\eta_Q$, parameters $\sigma$,$\rho$,$\tau$,$\gamma$, NCP $1-\beta$}
	\State{Initialize $2\cdot|\mathcal{A}|$ QR models’ and RL agent’s parameters with random weights}
	\For{$t$=1,2,...}
	    \State{Given the input state $\bm{x}_t$, with the probability $\epsilon$, the agent selects a random action; otherwise the agent selects $\alpha_t^*=\mathop{\arg\max}Q_\pi(\bm{x}_t,\bm{\alpha})$}
	    \Statex{\textit{// Update parameters for online quantile predictors}}
	    \State{Execute action in the environment: Select the QR models for LUB’s quantiles with probability proportions $\alpha_t^*$ and $\alpha_t^*+1-\beta$, and predict the corresponding quantiles: $\hat{q}_t^{\alpha_t^*}=f^{\alpha_t^*}(\bm{x}_t;\hat{\bm{W}}_{t-1}^{\alpha_t^*})$,$\hat{q}_t^{\alpha_t^*+1-\beta}=f^{\alpha_t^*+1-\beta}(\bm{x}_t;\hat{\bm{W}}_{t-1}^{\alpha_t^*+1-\beta})$}
	   \State{Reveal true value $y_t$}
	   \State{Respectively store the experience $e_t=(\bm{x}_t,y_t)$ in the selected predictors’ buffers $D_t^{\alpha_t^*}$,$D_t^{\alpha_t^*+1-\beta}$ with maximal priority $p_t^\alpha=\mathop{max}_{i<t}p_i^{\alpha_t^*}$, $p_t^{\alpha_t^*+1-\beta}=\mathop{max}_{i<t}p_i^{\alpha_t^*+1-\beta}$}
	   \State{Execute the \textit{Line 6 – Line 13} of Algorithm 1 to update QR models’ parameters for the selected LUB's predictors, respectively.}
	   \Statex{\textit{// Update parameters for RL’s agent}}
	   \State{Calculate the reward in (8) and update the input feature $\bm{x}_{t+1}$ for the next time-step based on the true load value $y_t$}
	   \State{Store the transition $(\bm{x}_t,r_t,\alpha_t^*,\bm{x}_{t+1})$ in the dueling DQN’s buffer $D_t^Q$}
	   \State{Sample random minibatch of transitions from $D_t^Q$, and update the agent’s network using (17)}
	\EndFor    
	\end{algorithmic}  
	
\end{algorithm}

\section{Case Study}

This section testifies the effectiveness of the proposed method based on real-world datasets. And the aim is to show the proposed online method (1) compared with online CPI method, adjusts adaptively to either symmetric or unsymmetric data distributions, (2) by learning from newly arriving data, achieves better performance than offline-trained PI estimation methods, and (3) compared with offline-trained CPI and OPI methods, is robust against the concept drift. Also, the effectiveness of the PER strategy is demonstrated in subsection \textit{E}, and convergence performance is analyzed in subsection \textit{F}.

\subsection{Implementation Details}
An hourly ahead forecasting is investigated. Smart meter data from Low Carbon London trail \cite{32} is used to verify the proposed method. In the trail, the customers receive the Time-of-Use (ToU) tariff are in the ToU group, while those receive the flat tariff are in the non-ToU group. To validate the proposed method is effective to data with different underlying distributions and to PIs with different NCP, the two cases are considered. And PIs with 95$\%$ and 90$\%$ NCP are constructed for the two cases respectively. The first case is based on the aggregated hourly load in the year of 2013 of 251 customers in the ToU group. The second case considers the net load, which is the aggregated hourly load of 251 customers in the ToU group minus the hourly wind output. The probability distributions of data in the two cases are different. Specifically, it is usually believed that the probability distribution of the load is approximated with the Normal distribution. Therefore, the distribution of the data in the first case has relatively small skewness. And the probability distribution of the wind is believed to be approximated with the Weibull distribution, which is positive skew with large skewness. Therefore, the distribution of the net load in the second case is positively skewed and has relatively larger skewness. 

\begin{table}[h]
\caption{Summary of Quantile Predictors’ Parameters}
\begin{center}
\begin{tabular}{ c  c }
\hline\hline
    Item & Value\\
\hline
    Batch size & 128\\
    No. of neurons in each hidden layer & 128\\
    No. of hidden layers & 1\\
    No. of neurons in input layer & 168\\
    No. of neurons in output layer & 1\\
    Optimizer & Adam\\
    Learning rate & 1e-3\\
\hline\hline
\end{tabular}
\end{center}
\end{table}

\begin{table}[h]
\caption{Summary of Dueling DQN’s Agent’s Parameters}
\begin{center}
\begin{tabular}{ c  c }
\hline\hline
    Item & Value\\
\hline
    Batch size & 128\\
    No. of hidden layers & 2\\
    No. of neurons in the first hidden layer & 512\\
    No. of neurons in the second hidden layer & 256\\
    No. of neurons in input layer & 168\\
    Optimizer & Adam\\
    Learning rate & 1e-4\\
\hline\hline
\end{tabular}
\end{center}
\end{table}

Since the specific forecasting model is not the main focus of the work, quantile multi-layer perception (QMLP) is used for the 2$|\mathcal{A}|$ quantile predictors.  There is no restriction to the type of machine learning-based predictor in the proposed framework. As long as the predictor can incrementally update its model when new data is collected, it can be used in the proposed setting. For example, NN can update parameters from online experiences fairly well, so many other types of NN-based models can be used here. The historical load 168 hours ago is used as the input for both predictors’ feature and agent’s state. The quantile predictors’ parameters are summarized in Table \uppercase\expandafter{\romannumeral1}, along with the parameters of Dueling DQN’s agent summarized in Table \uppercase\expandafter{\romannumeral2}, where the number of neurons in output layer is equal to the number of actions which will be discussed in the later subsection. Winkler, average coverage deviation (the reliability/calibration score), and sharpness scores are used as the three criteria \cite{18} for comparison. The program is implemented on the laptop with Intel®CoreTM i5-10210U 1.6 GHz CPU, and 8.00 GM RAM\footnote{\textcolor{blue}{Codes will be available at https://github.com/YufanChang/Optimal-Adaptive-Prediciton-Interval after publication.}}. 

\subsection{Comparison with Online CPI Forecasting Methods}
Under the cases of load and net load, the proposed online closed-loop OPI forecasting with PER is compared with online CPI forecasting with PER. The number of quantile proportions changes from 3 to 63. The results are shown in Fig. 4, where the results of the load are in Fig. 4 (a), and the results of the net load are in Fig. 4 (b). The Winkler score is used as the overall quality criteria of PIs, which is calculated as the average values over the entire forecasting steps. 

\begin{figure}[h]
  \centering
  \includegraphics[scale=0.45]{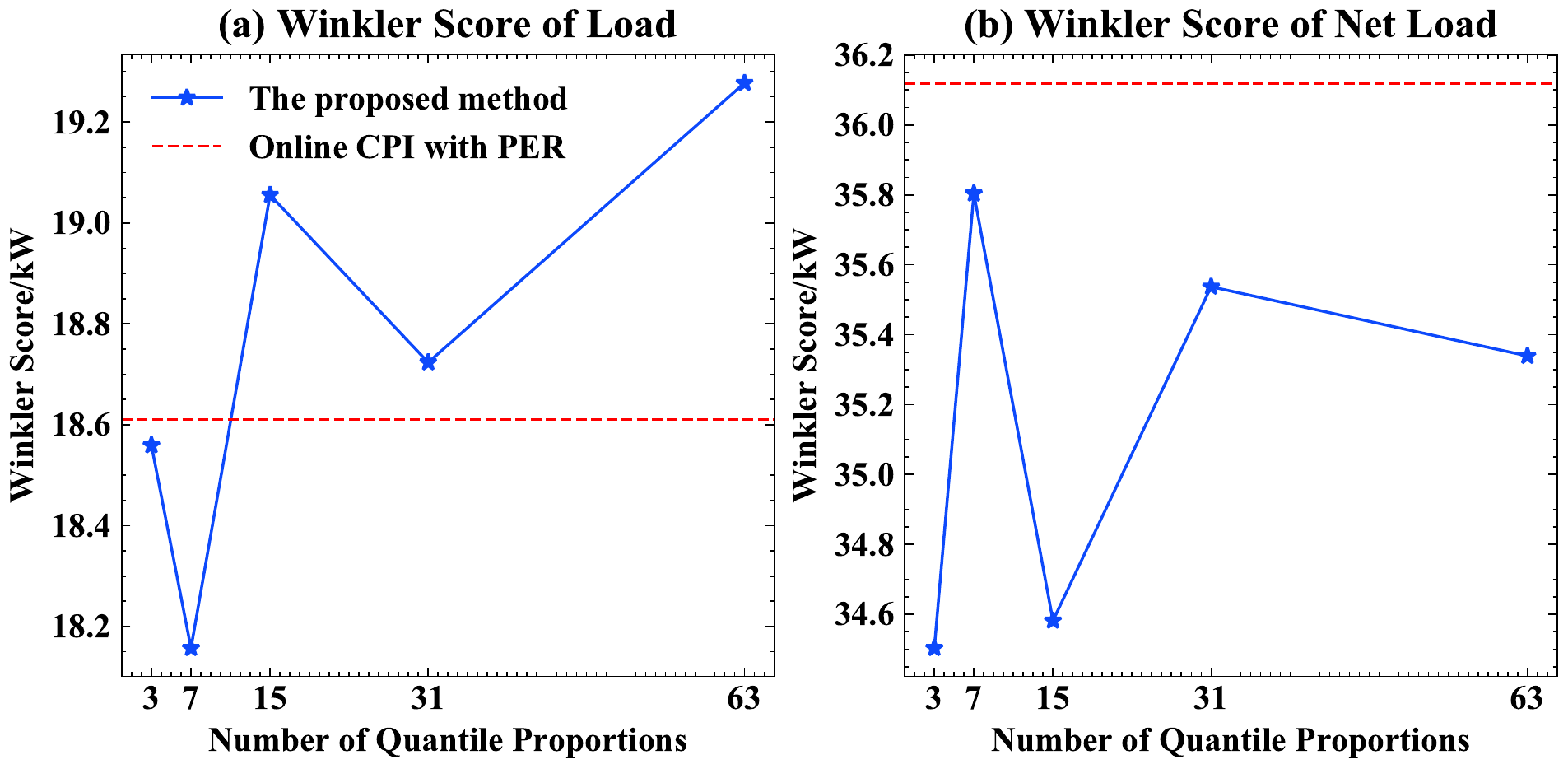}\\
  \caption{Comparison of Winkler scores of the proposed method and online CPI in two cases.}
\end{figure}

In terms of the Winkler score, the proposed method and online CPI forecasting with PER show interesting results under the cases of load and net load. With the change of the number of nominal proportions, in the case of the load, the Winkler scores of the proposed method fluctuate around the Winkler score of online CPI forecasting with PER; in the case of the net load, the Winkler scores of the proposed method are lower than that of online CPI forecasting with PER, which indicates the better performance. This is because the data in the first case has small skewness, the optimal lower and upper quantiles are approximated with that of the CPI. Therefore, the proposed method and online CPI forecasting with PER have the similar performances. However, for the second case, with the penetration of the wind, the distribution of the net load becomes more complex and has larger skewness. The optimal lower and upper quantiles are very different from that of the CPI. Thus, online CPI forecasting with PER cannot adapt to the distribution of the net load, which results in the poor performance. Also, we notice that there are some fluctuations in results of Fig. 4 under different sizes of action space. It is brought by the exploration mechanism of the RL algorithm and uncertainty caused by quantile crossing, a common concern in QR models. With the increase penetration of the wind, the skewness of the net load distribution becomes more evident. Thus, the online CPI forecasting with PER becomes more unsuitable. To prove this point, the noises sampled from the Beta distribution $4\cdot Beta(\alpha=2,\beta=5)$ are added to the net load to increase the skewness. The relative Winkler scores, which is the Winkler score of online CPI forecasting with PER minus the scores of the proposed method, are shown in Fig. 5. It shows that the improvements of the Winkler scores become more obvious when added with the noise. So the results demonstrate the effectiveness of the proposed method which adaptively selects optimal quantiles conditioned on the data distribution.

\begin{figure}[h]
  \centering
  \includegraphics[scale=0.7]{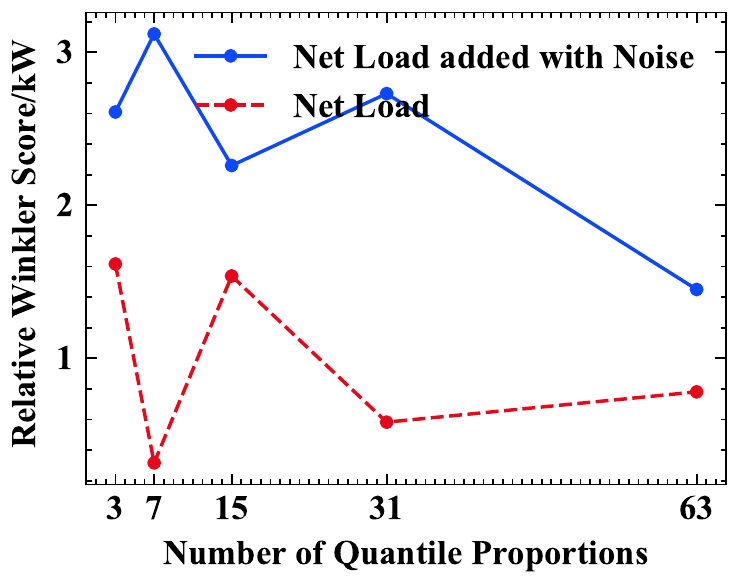}\\
  \caption{Relative Winkler score of the net load under different levels of noise.}
\end{figure}

\subsection{Comparison with Offline-trained PI Forecasting Methods on Test Set}
To further evaluate the forecast performance of the proposed approach, offline-trained PI forecasting comparison candidates belonging to the categories of CPI and OPI forecasting approaches are investigated. Specifically, the mixed integer programming model (MLMIP) \cite{18}, belonging to adaptive OPI methods, is used. And the CPI methods include the light quantile gradient boosting regression tree (Light QGBRT), QMLP guided by the pinball loss, and a naive benchmark that is an extension of persistence model in probabilistic setting, which constructs PIs for each hour based on symmetric quantiles derived from historical statistical data \cite{8}. Here, the network parameters of QMLP method are the same as that of the quantile predictors of the proposed method. To make fair comparison, although the proposed online method is performed in a stream fashion and there is no need of the dataset division, the data is partitioned into the training and test sets of size 70$\%$ and 30$\%$ for the comparison. The numbers of quantile proportions of the proposed method are chosen as 7 and 3, respectively for the two cases. For the proposed method and comparison candidates, the results on the test set after the learning on training set are presented in Fig. 6. 

\begin{figure}[h]
  \centering
  \includegraphics[scale=0.5]{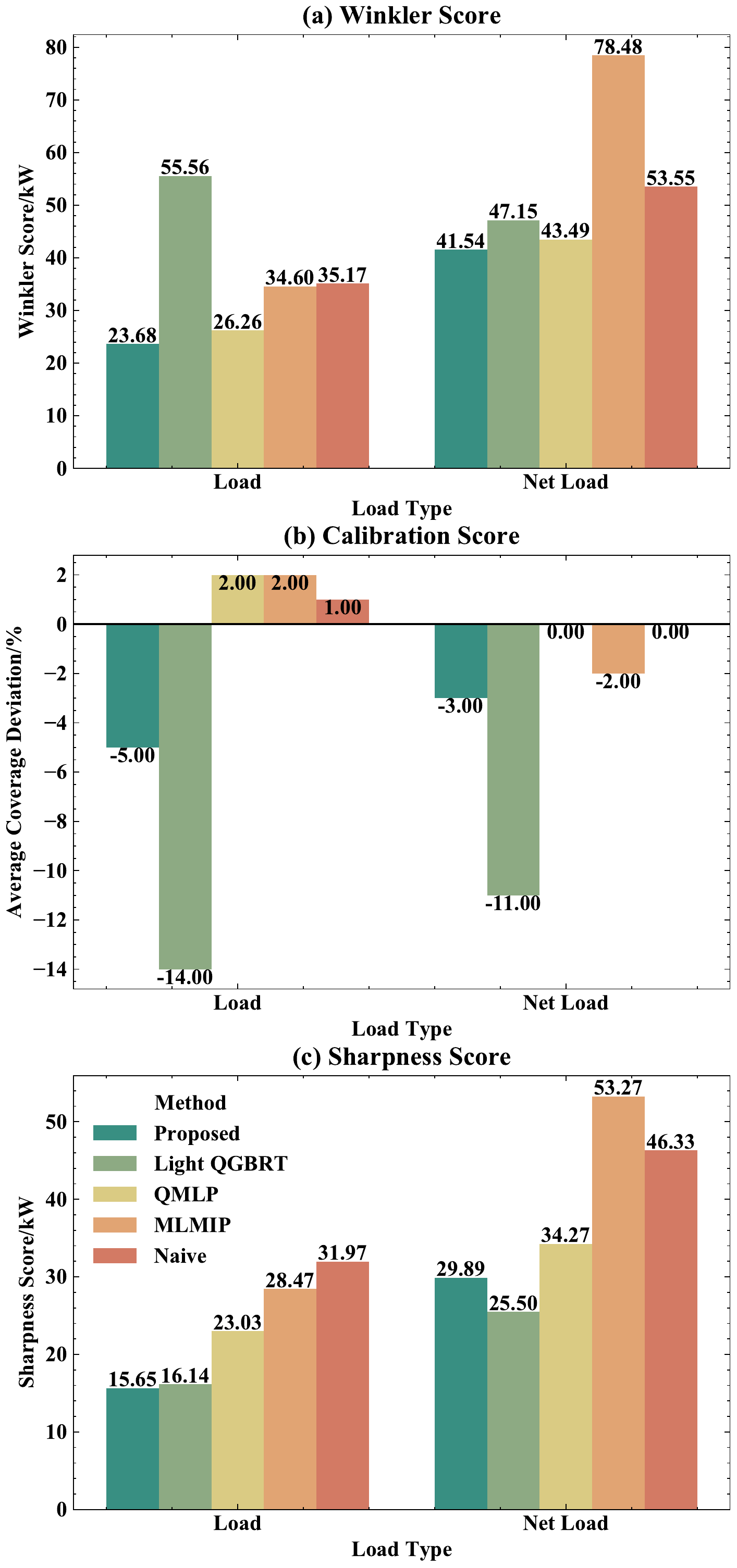}\\
  \caption{Quality criteria of PIs of the proposed method and offline comparison candidates on test set in the two cases.}
\end{figure}

Due to the constraint of restricting the empirical coverage probability no less than the NCP, MLMIP method always has good calibration performance. However, the Winkler score of MLMIP method is not satisfying, especially for the PI forecasting task in the second case. It is inferred that although MLMIP method is independent of distribution assumption, and demonstrates its good performance for the load level in MW, the relatively simple forecasting model, ELM that the approach relies on, may dampen its ability to capture the complex data distribution that the net load has, which is more variable and uncertain. Also, if like \cite{18}, multiple MLMIP models are trained for each month separately, the performance of it is expected to be improved. Since the comparisons in this subsection focus on the performance of one particular model for the whole dataset, this case is not investigated.

In terms of Winkler score, in the two cases, the proposed method has the lowest Winkler score on the test set. Compared with the naive benchmark, it improves the performance by 33$\%$ in the first case, and by 22$\%$ in the second case. Among the offline methods, QMLP performs the best. However, the QMLP-based method sacrifices the sharpness for the high reliability. Specifically, in the first case, although the CPI is approximated with the OPI, the proposed method’s merit, i.e., consecutive learning from the new data, enables it to better capture the future trend of the load. Therefore, the proposed method has the smaller Winkler score with sharper PIs than the QMLP-based method. In the second case, the improvement of the Winkler score compared with the best comparison candidate is mainly due to the CPI forecasting method’s incapability of adapting to data with large skewness. 

Fig. 7 illustrates the 168 hours PIs on test set predicted by the proposed method, superimposed with the CPIs predicted by the QMLP in the two cases. In Fig. 7 (a), the proposed method achieves the lower upper bounds and higher lower bounds than QMLP. When the load displays large variability, the width of PI is relatively large to capture the uncertainty. During the other periods, the upper and lower bounds of the proposed method are very close to the true load value. The good performance can be attributed to its adaptive adjustment to the stream of data. In Fig. 7 (b), the net load displays the larger uncertainty than the load in Fig. 7 (a). And the negative value of the net load means the power flows from the demand side to the grid. The width of PI produced by QMLP is relatively larger than the proposed method. And the proposed method adapts well to the sudden decrease of the net load, which is partly caused by the sudden increase of the wind power. 

\begin{figure}[h]
  \centering
  \includegraphics[scale=0.45]{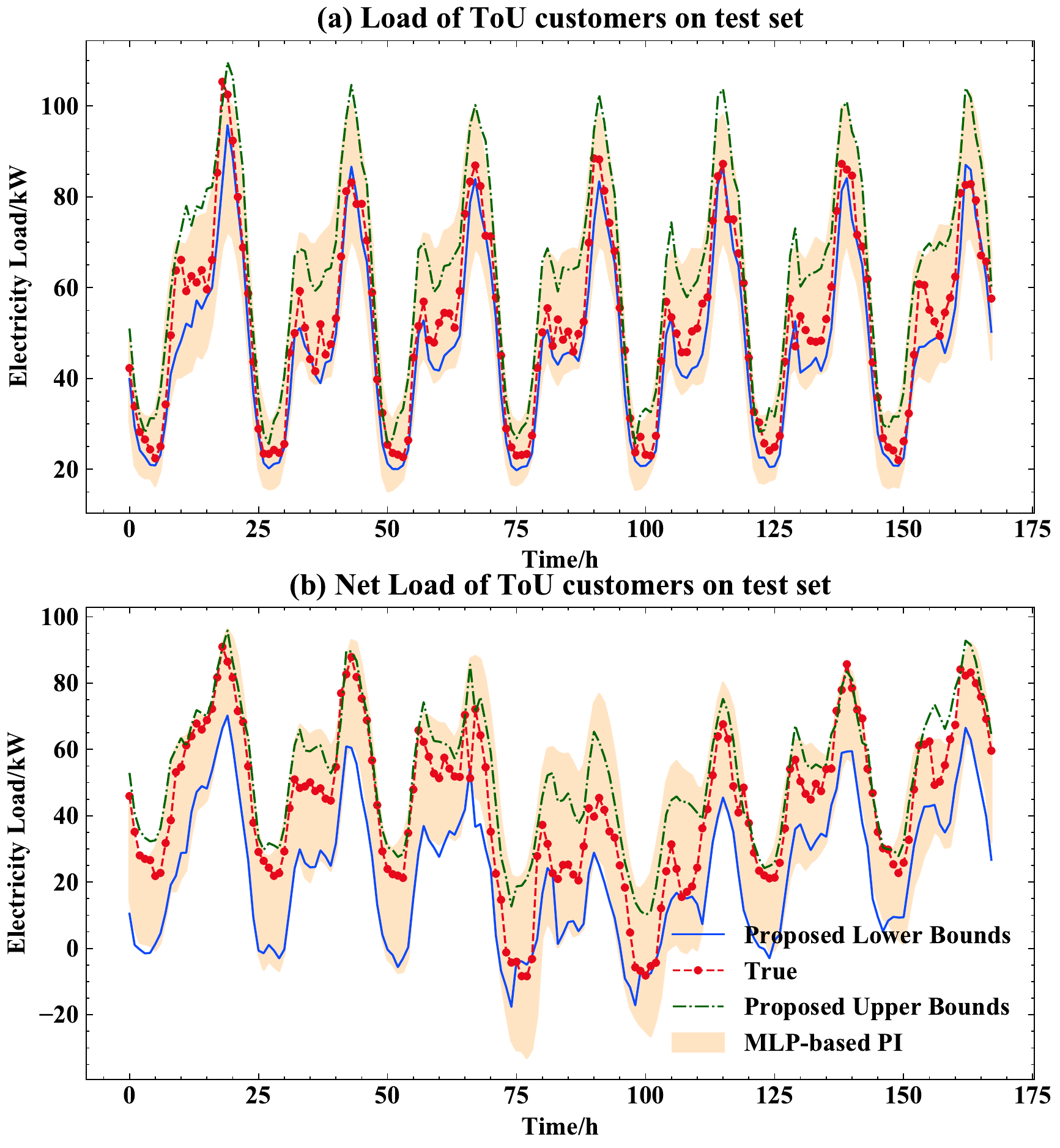}\\
  \caption{168 hours PIs forecast on test set obtained via the proposed method and QMLP in two cases.}
\end{figure}

The time for updating the model parameters of the proposed method and offline-trained comparison candidates is shown in Table \uppercase\expandafter{\romannumeral3}. Since the updating time for the load and net load cases is similar, only the time in the net load case is displayed. It shows that the proposed method takes the smallest time for updating the model parameters. Among the offline-trained methods, since Light QGBRT is a light model designed to be computationally efficient, it takes the smallest time. The results in Table \uppercase\expandafter{\romannumeral3} are reasonable. The model updating of the proposed method only relies on a small batch of data. In contrast, the model updating of the offline methods involves the large numbers of iterations, and the model can only be updated until enough samples are collected. 

\begin{table}[h]
\caption{Comparison of time for updating the model parameters}
\begin{center}
\begin{tabular}{c  c  c  c}
\hline\hline
    The proposed method & Light QGBRT & QMLP & MLMIP\\
\hline
    96.8ms & 930ms & 21.5s & 1min 44s\\

\hline\hline
\end{tabular}
\end{center}
\end{table}

\subsection{Performance Comparison under the Concept Drift}
The concept drift means the statistical properties of the target variable, which the model is trying to predict, changes in unforeseen ways. To prove that the proposed online method is robust against the concept drift, the methods are examined on the scenarios where the training and test sets have different data distributions. Concretely, for the first case, the training set is replaced with the synchronous aggregated hourly load in the year of 2013 of 150 customers in the non-ToU group. For the second case, the training set is replaced with the same aggregated load in the non-ToU group minus the hourly wind output. And the test sets of the two cases remain the same as those in the subsection \textit{C}. Therefore, for the offline-trained comparison candidates, the results presented in the subsection \textit{C} are regarded as the benchmarks to show the results under no concept drift. For the proposed method, the performances after online learning on the same test sets under no concept drift are the benchmarks.

The Winkler scores on the test set are shown in Fig. 8. In the two cases, the proposed method has the similar Winkler scores with the benchmarks, and has the best performance among the comparison candidates. Thanks to the PER strategy, the data following the new distribution are more likely to be sampled frequently for the QR models to learn. Therefore, the method can efficiently adapt to the changes in load patterns. For the offline-trained comparison candidates, the concept drift deteriorates the performances. Especially in the second case where the data distribution is more complex, the offline-trained methods have larger Winkler score under the concept drift. For example, the Winkler scores of MLMIP, Light QGBRT, and Naive method are nearly twice as large as the benchmarks. Therefore, the results solidly verify the better performance of the proposed method.
\begin{figure}
  \centering
  \includegraphics[scale=0.4]{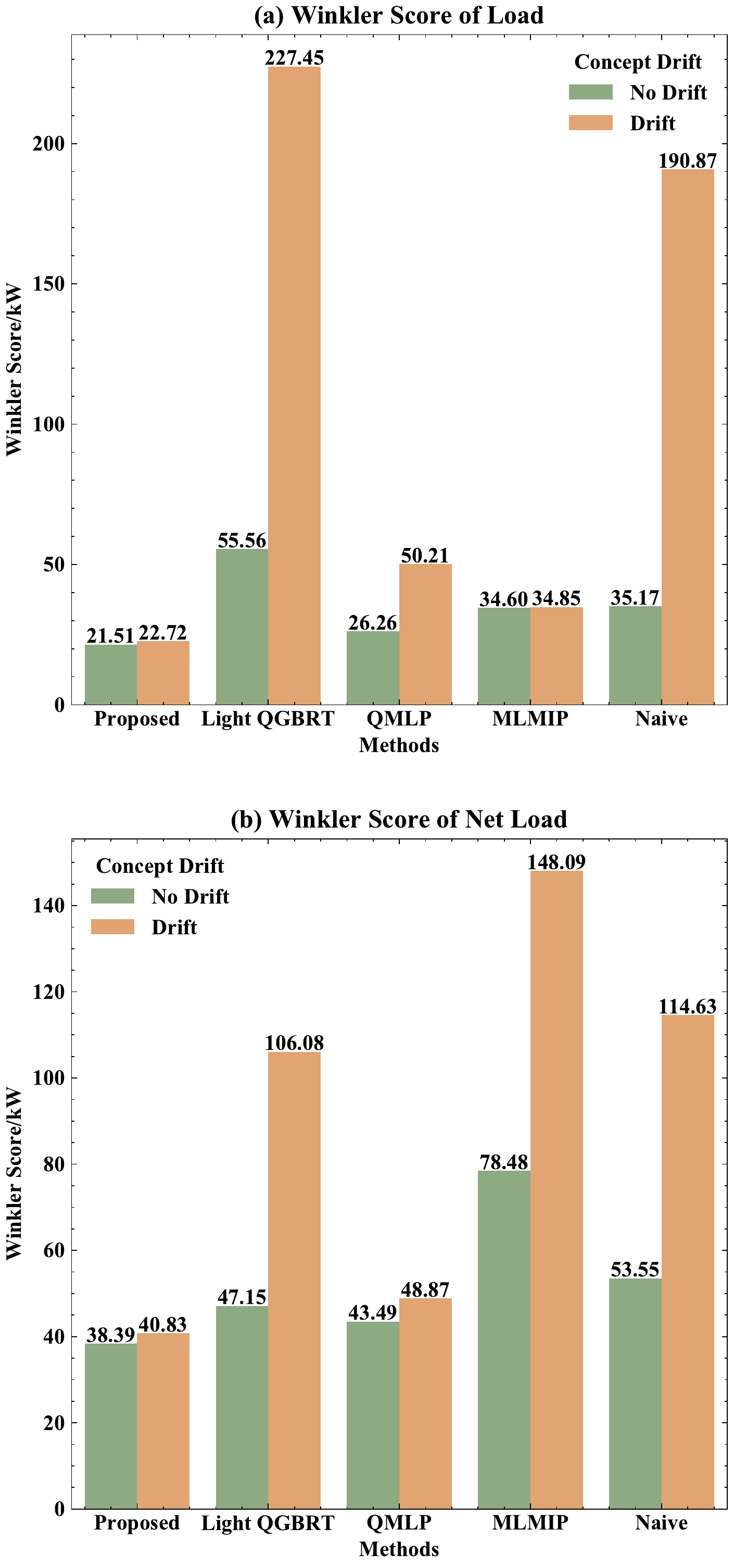}\\
  \caption{Winkler score of PIs of the proposed method and offline comparison candidates on test set in the two cases under concept drift.}
\end{figure}

\subsection{Comparison with Online OPI Forecasting Without PER Strategy}
To demonstrate the effectiveness of PER strategy, the proposed method is compared with its counterpart without PER strategy, where a small batch of data are randomly sampled from buffer to update the QR model recursively. The results over the entire forecasting steps are shown in Fig. 9, where the results of the load are in the left column, and the results of the net load are in the right column.

\begin{figure}[h]
  \centering
  \includegraphics[scale=0.37]{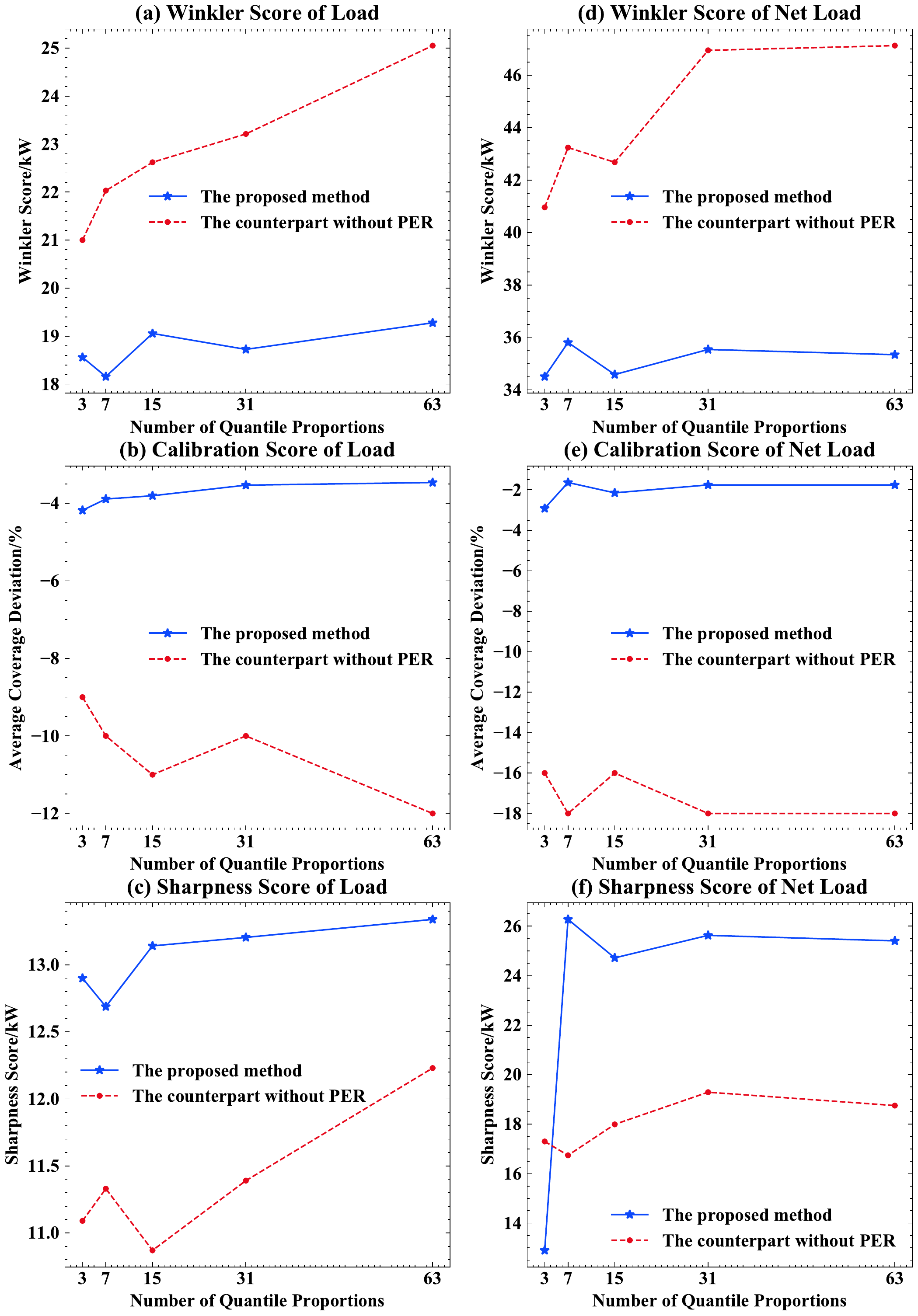}\\
  \caption{Quality criteria of PIs of the proposed method and its counterpart without PER strategy in the two cases with the change of quantile proportions.}
\end{figure}

It is evident that in the two cases, the counterpart without PER strategy has the most satisfactory sharpness performance. However, the calibration of this approach undergoes the significant discrepancy from the NCP. Since the sharpness and calibration only display the one aspect of the PI quality, the overall metric such as Winkler score is needed for the performance evaluation. In terms of the performance evaluated by the Winkler score, the proposed method outperforms the online counterpart without PER strategy. The result demonstrates the effectiveness of the PER technique. Therefore, the ways that the experiences are sampled to update predictors’ parameters can affect the performance of QR models for online quantile forecasting. Compared with updating the predictors’ parameters by random experiences in the buffer, PER enables the learning to be more effective, because the samples hard to predict and the new sample are learned more frequently by the QR models.

\subsection{Convergence Analysis}
To demonstrate that the proposed method has good convergence performance, we show PIs in the first 168 hours of QR models’ learning processes in Fig. 10, and the moving average rewards gained by agent in Fig. 11. It is shown that the upper and lower quantiles produced by QR models cannot capture the trends in load or net load profiles at the very beginning, but QR models learn quickly to issue accurate predictions in a few steps and obtain PIs with good quality. For agent’s policy learning, the achieved average rewards increase during the learning process, and converge soon to the maximum value. The learning processes of QR models and RL’s agent can affect each other. However, since QR models and RL’s agent work toward the same goal, namely improving PI’s quality, the dynamic learning processes of QR models and RL’s agent are stable and converge fast.

\begin{figure}[h]
  \centering
  \includegraphics[scale=0.45]{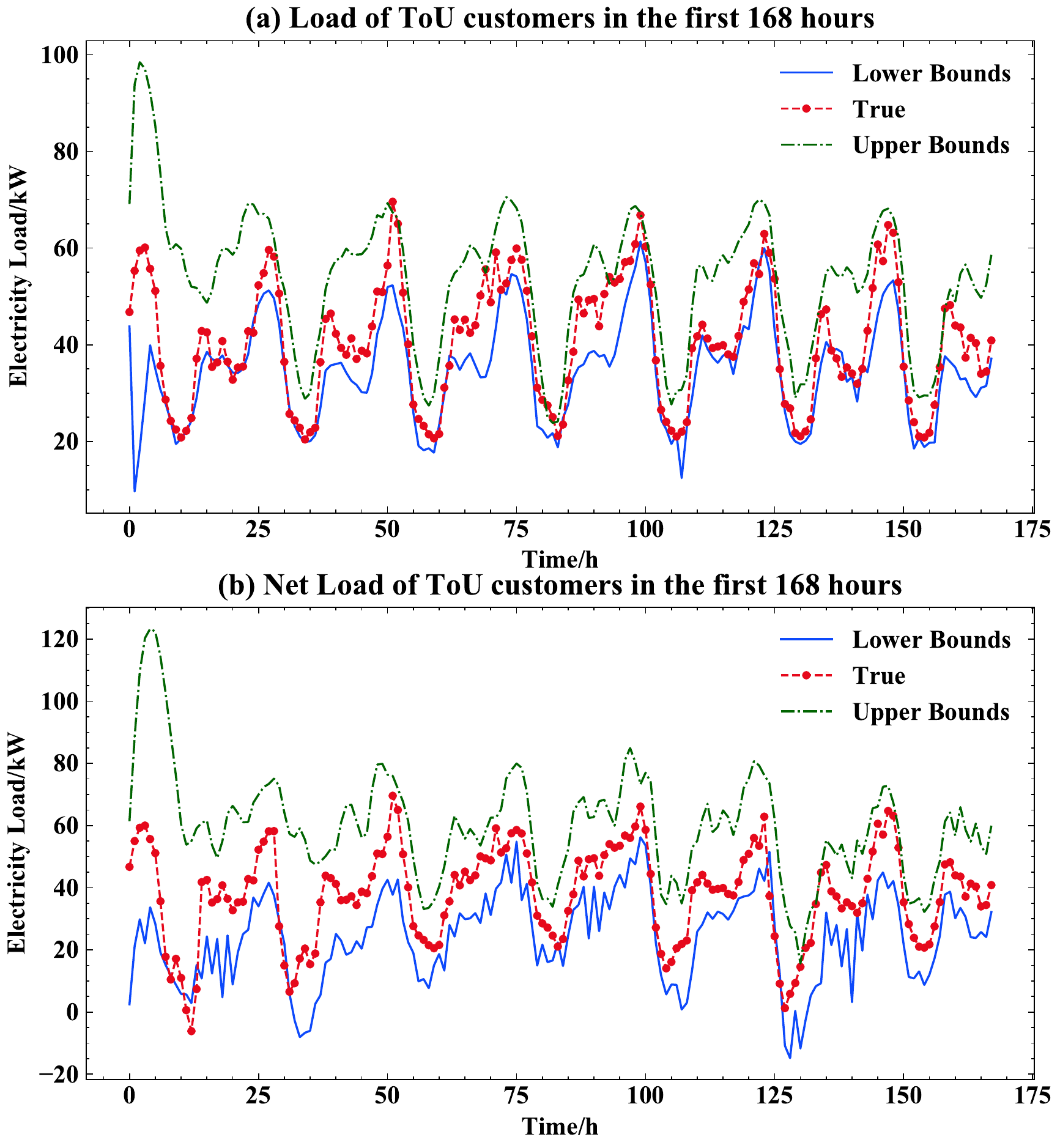}\\
  \caption{PIs in the first 168 hours of QR models’ learning processes.}
\end{figure}

\begin{figure}[h]
  \centering
  \includegraphics[scale=0.8]{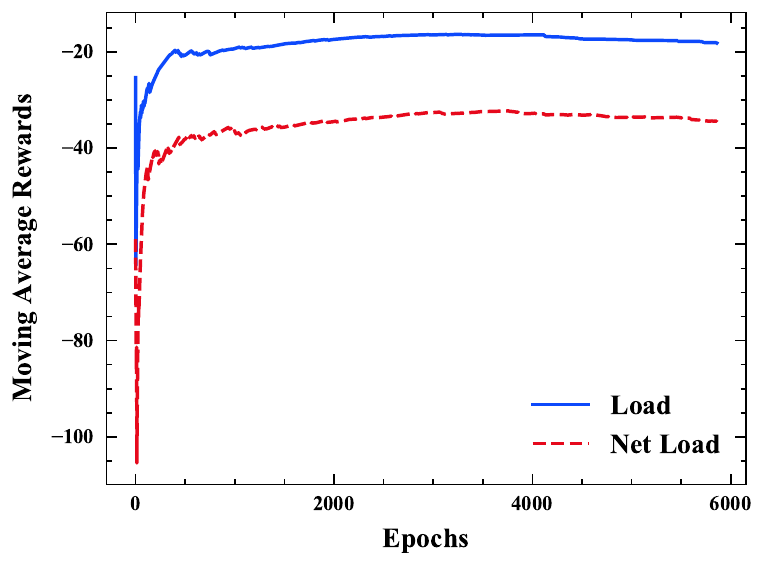}\\
  \caption{RL’s agent’s moving average rewards during the learning process.}
\end{figure}

\section{Conclusions}
In this paper, we propose a new approach to determine the optimal adaptive prediction intervals for load forecasting in distribution networks via reinforcement learning. It provides optimal prediction intervals adaptive to different data distributions. The approach uses the RL method to unify two correlated sub-problems, i.e., probability proportion selection and corresponding quantile forecasting, which allows it to be implemented in an online closed-loop fashion. Case studies on both load and net load forecasting have shown that the proposed method can produce good quality PIs. Compared with the online CPI method, it selects the optimal quantiles’ probability proportions with good flexibility, and thus adapts well to the skewness of data distributions. Also, by learning from the load incrementally, it obtains lower Winkler score and is more robust against concept drift compared with offline-trained methods. The case studies also reveal that the PER strategy improves the online learning efficiency, and helps to achieve better performance. 

 In this work, we approximate the probability proportions in a discrete action space. Therefore, it is still required to consider representing probability proportions in a continuous action space in the future. Besides, it is worthy to note that the proposed closed-loop framework is a general way that links two separate tasks working toward the same goal. Its potential application is not restricted to forecasting problems in power systems; instead, it may be valuable to investigate the applications of the closed-loop framework on other problems consisting of two separate tasks or stages.

\section*{Acknowledgement}
This work was supported by the National Natural Science Foundation of China (U1866206).

\bibliographystyle{IEEEtran}
\bibliography{IEEEabrv,mylib}
\end{document}